\documentclass[letterpaper,twocolumn,english,prl,epsfig,superscriptaddress,showpacs]{revtex4-1}
\usepackage[T1]{fontenc}
\usepackage{amsmath}
\usepackage{graphicx}
\usepackage{amssymb}


\usepackage{babel}

\begin{document}

\title{Theory of spike timing based neural classifiers}

\author{Ran Rubin}

\affiliation{Racah Institute of Physics, Hebrew University, 91904 Jerusalem, Israel}

\affiliation{Laboratoire de Physique Th\'{e}orique de l'ENS, CNRS, Univ. Paris
6, 24 rue Lhomond, 75005 Paris, France}

\author{R\'{e}mi Monasson}

\affiliation{Laboratoire de Physique Th\'{e}orique de l'ENS, CNRS, Univ. Paris
6, 24 rue Lhomond, 75005 Paris, France}

\affiliation{Simons Center for Systems Biology, Institute for Advanced Study,
Einstein Drive, Princeton, NJ 08540, USA}

\author{Haim Sompolinsky}

\affiliation{Racah Institute of Physics, Hebrew University, 91904 Jerusalem, Israel}

\affiliation{Interdisciplinary Center for Neural Computation, Hebrew University,
91904 Jerusalem, Israel}

\affiliation{Center for Brain Science, Harvard University, Cambridge, Massachusetts
02138, USA}
\begin{abstract}
We study the computational capacity of a model neuron, the Tempotron,
which classifies sequences of spikes by linear-threshold operations.
We use statistical mechanics and extreme value theory to derive the
capacity of the system in random classification tasks. In contrast
to its static analog, the Perceptron, the Tempotron's solutions space
consists of a large number of small clusters of weight vectors. The
capacity of the system per synapse is finite in the large size limit
and weakly diverges with the stimulus duration relative to the membrane
and synaptic time constants.
\end{abstract}


\maketitle
\begin{figure}[t]
 \includegraphics[width=1\columnwidth,height=0.4\columnwidth]{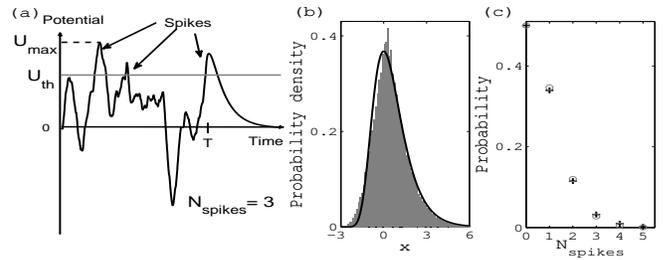}
\caption{\label{fig:traces-1} (a) Example of voltage traces $U(t)$. (b) Probability
density of the rescaled maximal potential $x$ as defined in eq. \eqref{eq:v_max}
with a fitted scale factor $\beta$. (c) Probability of $N_{\text{spikes}}$.
The Line in (b) is a standard Gumbel law. In (c) circles indicate
the theoretical Poisson law. Data was measured with $K=400,\ \alpha=1.68,\ N=500$
and 34 samples.}

\end{figure}

Neural network models of supervised learning are usually concerned
with processing static spatial patterns of intensities. A famous example
is learning in a single-layer binary neuron, the Perceptron \citep{minsky1988pee,gardner1987msc}.
However, in most neuronal systems, neural activities are in the form
of time series of spikes. Furthermore, stimulus representation in
some sensory systems are characterized by a small number of precisely
timed spikes \citep{johansson2004first,gollisch2008rapid}, suggesting
that the brain possesses a machinery for extracting information embedded
in the timings of spikes, not only in their overall rate. Thus, understanding
the power and limitations of spike-timing based computation and learning
is of fundamental importance in computational neuroscience. G\"{u}tig
and Sompolinsky \citep{gutig2006tnl} have recently suggested a simple
model, the Tempotron, for decoding information embedded in spatio-temporal
spike patterns. The Tempotron is an Integrate and Fire (IF) neuron,
with $N$ input synapses of strength $\omega_{i}$, $i=1,\dots,N$.
Each input pattern is represented by $N$ sequences of spikes, where
the spike timings for the afferent $i$ are denoted by $\{t_{i}\}$.
The membrane potential is given by \begin{equation}
U(t)=\sum_{i=1}^{N}\omega_{i}\sum_{t_{i}<t}u(t-t_{i})\label{eq:Tempotron_voltage-1}\end{equation}
 where $u(t)$ denotes a fixed causal temporal kernel. An example
is the difference of exponentials form: $u\left(t\right)=u_{0}\left(e^{-\frac{t}{\tau_{m}}}-e^{-\frac{t}{\tau_{s}}}\right)$,
where $\tau_{m}$ and $\tau_{s}$ correspond, respectively, to the
membrane and synaptic time constants %
\footnote{In all the numerical results presented here we have used $\tau_{s}=\tau_{m}/4$
except in Fig. 2b%
}. The Tempotron fires a spike whenever $U$ crosses the threshold,
$U_{\text{th}}$, from below %
\footnote{\label{fn:fn1-1}In this work effects of potential reset after a spike
are not relevant%
} (Fig. \ref{fig:traces-1}a). The Tempotron performs a binary classification
of its input patterns by firing one or more output spikes when presented
with a 'target' (+1) pattern and remaining quiescent during a 'null'
(-1) pattern. %

In this Letter we present a theoretical study of the computational
power of the Tempotron. We focus on the standard task of classifying
a batch of $P=\alpha N$ random patterns, where $\alpha$ denotes
the number of patterns per input synapse. For each pattern, the timings
of the input spikes from each input neuron are randomly chosen from
independent Poisson processes with rate $\frac{1}{T}$, where $T$
is the duration of the input patterns, and the desired output, $y=\pm1$,
is randomly and independently chosen with equal probabilities. A solution
to the classification problem is a set of synaptic weights $\{\omega_{i}\}$
that yields a correct classification of all $P$ patterns. We will
address several fundamental questions. First, numerical simulations
based on a simple error-correcting on-line learning algorithm suggest
that the capacity of the IF neuron namely, the maximal number of patterns
per synapse, $\alpha_{c}$, which, with high probability (approaching
1 for large $N$), can be correctly classified is independent of the
number of input synapses \citep{gutig2006tnl}; however, an analytical
proof for this property has been lacking. Secondly, it is important
to understand how the computational capabilities of the neuron depend
on the various time scales in the dynamics of the system. Finally,
our study highlights the complex geometric structure of the space
of solutions for $\alpha<\alpha_{c}$, similar to the one arising
in other hard computational problems, such as learning in multilayered
neural networks \citep{engel2001sml} or random combinatorial optimization
\citep{Monasson20071,mezard2009information}.

\begin{figure}
 \includegraphics[width=1\columnwidth,height=0.4\columnwidth]{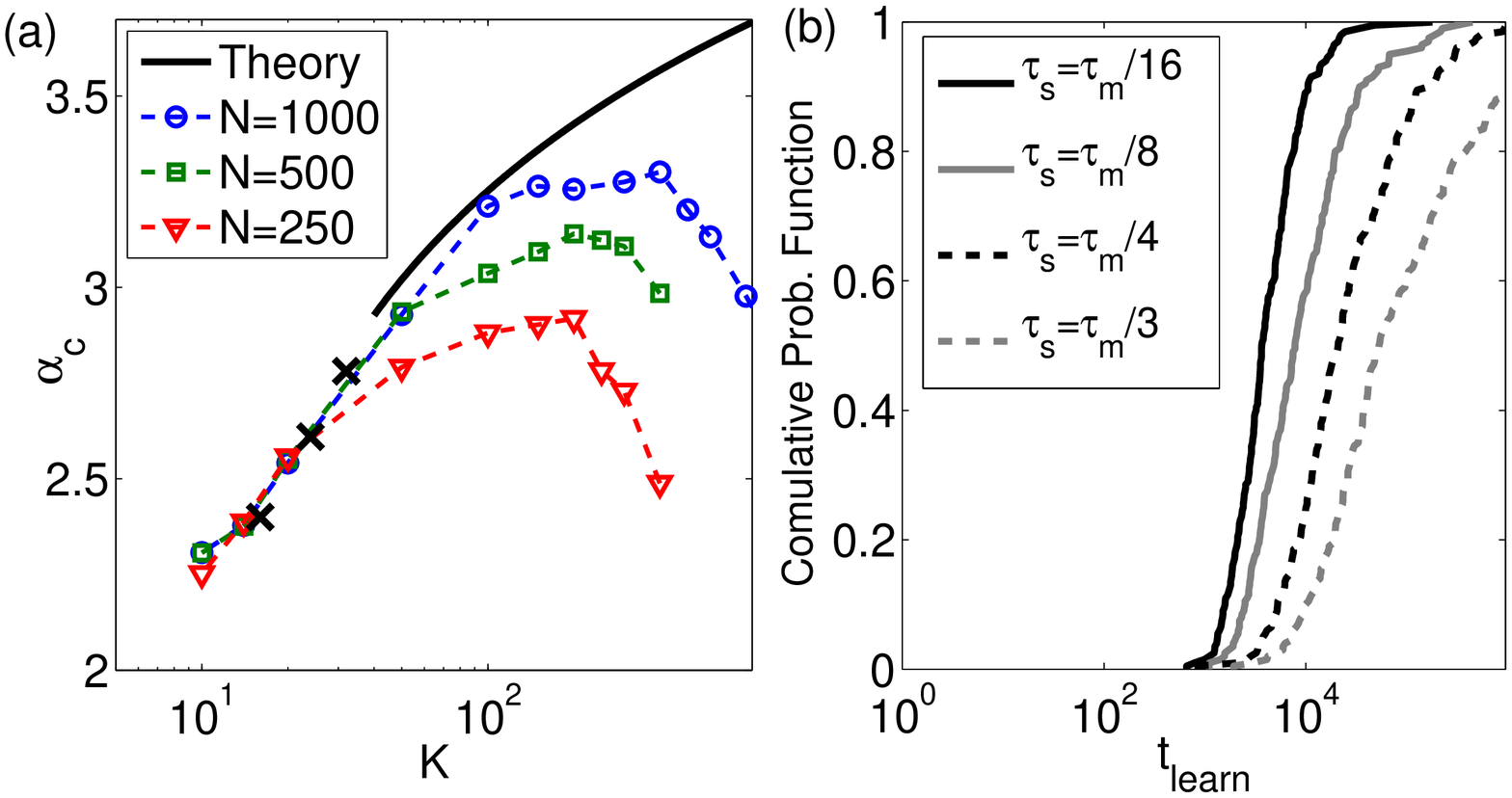}
\caption{\label{fig:capacity} (a) Capacity $\alpha_{c}$ of the Tempotron
vs. $K$. Lines with symbols show results from the learning algorithm
of \citep{gutig2006tnl}. The solid line shows the large--$K$ theory
\eqref{eq:alpha_c}, with an additive constant ($\alpha_{c}=\left(\ln\ln K\right)/2\ln2+\alpha_{0}$,
with $\alpha_{0}=2.58$) fitted to the predictions of the replica
method for the discrete Tempotron for $K^{\text{discrete}}=2,3,4$
($\times$ symbols). To compare the theory of the discrete Tempotron
with the simulation results of the continuous time Tempotron we used
$K^{\text{discrete}}=K/8$. (b) Distribution of learning times for
different $\tau_{s}$, and for fixed $\tau_{m}=T/25$, $\alpha=2.6$
and $N=1000$. As $\tau_{s}$ decreases so does the mean learning
time, indicating that |$\alpha-\alpha_{c}|$ has increased, as predicted
by eqs. \eqref{eq:K} and \eqref{eq:alpha_c}. }

\end{figure}

Our theoretical analysis, presented below, shows that a fundamental
parameter is the pattern duration, $T$, relative to the neural time
scales, \begin{equation}
K=\frac{T}{\sqrt{\tau_{s}\tau_{m}}}\ .\label{eq:K}\end{equation}
 The properties of the Tempotron can be most easily understood, when
both $N$ and $K$ are large, with $N\gg K$. This limit is biologically
sensible if we consider a neuron with $N\sim10^{3}$ synapses, inputs
that are presented for $T\sim100-1000$ milliseconds, and constants
$\tau_{s}\sim1-10$, $\tau_{m}\sim10-100$ milliseconds. We predict
that, for any fixed $K$, the capacity is independent of $N$ in the
large $N$ limit. Furthermore, the capacity grows with $K$ as \begin{equation}
\alpha_{c}=\frac{\ln\ln K}{2\ln2}\ .\label{eq:alpha_c}\end{equation}
The convergence of the capacity to this expression is slow, requiring
that $\sqrt{\ln K}\gg1$. Nevertheless, this result has several qualitative
implications. Equation \eqref{eq:alpha_c} implies that the capacity
of the Tempotron is not bounded as $K$ increases, and may exceed
the capacity of the well-known Perceptron model ($\alpha_{c}=2$ \citep{gardner1987msc})
whose architecture is similar to the Tempotron. Note that when $K$
is $O\left(N\right)$, the few input spikes that arrive within a single
decision time window, $T/K$, do not carry sufficient information
to classify the patterns. We therefore expect that for any fixed $N$,
$\alpha_{c}$ is a non monotonic function of $K$ while the value
of $K$ that maximizes the capacity increases with $N$, as implied
by \eqref{eq:alpha_c}. This prediction is corroborated by numerical
simulations in Fig.~\ref{fig:capacity}a. Interestingly, according
to eq. \eqref{eq:K}, the performance should be sensitive also to
the short time behavior of the kernel as confirmed by the simulations
of Fig. \ref{fig:capacity}b. This short time behavior determines
how fast can the membrane potential change significantly. The faster
this cange can be, the easier it is to distinguish between inputs
that arrive within a short interval of time.

\begin{figure}[t]
 \includegraphics[width=1\columnwidth,height=0.4\columnwidth]{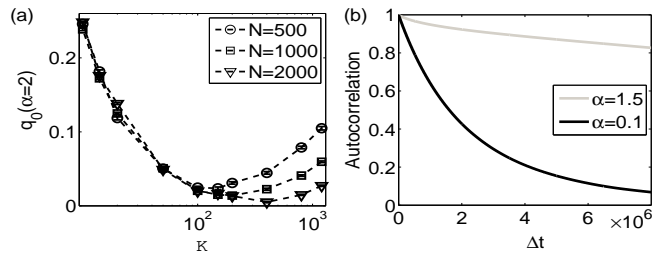}
\caption{\label{fig:OP}(a) Overlap between randomly chosen solutions, $q_{0}$,
at $\alpha=2$ as a function of $K$. (b) Auto-correlation function
$q_{AC}(\Delta t)\equiv\langle\hat{\omega}(t)\cdot\hat{\omega}(t+\Delta t)\rangle$
of a random walk inside a connected volume in solutions space for
$K=150$ and $N=500$. }

\end{figure}

In the Perceptron model, the solution space for a given classification
task is a convex volume, which shrinks in size and ultimately vanishes
as $\alpha$ approaches the capacity, $\alpha_{c}$. The overlap between
two typical solutions, $q_{0}$, defined by the inner product between
their normalized weight vectors, approaches $1$ at the critical capacity
\citep{gardner1987msc}. Our theory reveals that the solution space
of the Tempotron is of a strikingly different nature. First, the overlap
between two Tempotron weight vectors that solve the random classification
problem, $q_{0}$, approaches zero in the $K\gg1$ limit, for every
$\alpha<\alpha_{c}$. Secondly, the solution space is connected for
small $\alpha$ only. For larger values of $\alpha$, still far below
capacity, the solution space breaks into a large number of small disconnected
clusters, spread across the entire weight space. The overlap between
solutions within the same cluster, $q_{1}$, is close to $1$, while
two randomly chosen solutions are likely to lie in different clusters
and have overlap $q_{0}\approx0$. Simulations making use of the learning
algorithm of \citep{gutig2006tnl} support this picture. The overlap
between two solutions obtained from two different initial weight vectors
 vanishes for all values of $\alpha$ (Fig.~\ref{fig:OP}a). To probe
the overlap between solutions in the same cluster, we performed a
random walk in solution space \citep{barkai1992bsm}, starting from
a solution found by the Tempotron learning algorithm and rejecting
the random walk step attempts if they lead to a weight vector that
is not a valid solution. The auto-correlation function of this random
walk drops exponentially fast to zero for small $\alpha$, indicating
that the solutions space is connected, and hardly decays for higher
$\alpha(<\alpha_{c})$, as expected for a clustered solution space,
(Fig.~\ref{fig:OP}b).

% EVT

The above results are surprising and counter-intuitive since they
imply that even close to capacity, IF neurons with very different
weights can perform exactly the same classification, whereas IF neurons
with high degree of similarity in their weight vectors will typically
fail to solve the same task. To understand these properties we consider
a Tempotron whose $N$ weights are random variables drawn from any
probability distribution with finite first two moments. With no loss
of generality we may choose the mean and variance of the weights to
ensure that $U(t)$ has zero mean and unit variance. The threshold
potential, $U_{\text{th}}$, is such that a random pattern is classified
by each Tempotron as $\pm1$ with equal probabilities, \emph{i.e.,}
$U_{\text{th}}$ is the median value of the distribution of the maximum
of $U(t)$ over time, $U_{\text{max}}$. The synaptic potential $U(t)$
induced by a random input pattern approaches, in the large $N$ limit,
a temporally correlated Gaussian distribution. We use extreme value
theory (EVT) of Gaussian processes to evaluate the statistics of $U_{\text{max}}$
\citep{leadbetter1983extremes}. According to EVT, $U_{\text{max}}$
can be written as

\begin{equation}
U_{\text{max}}=U_{\text{th}}+\beta(x+\ln\ln{2})\label{eq:v_max}\end{equation}
where $x$ obeys the Gumbel density distribution $G(x)=\exp\left(-x-\exp(-x)\right)$,
whose median is $-\ln\ln2$. The scale factor is $\beta=1/\sqrt{2\ln{K}}+O\left(1/\ln{K}\right)$
and the threshold is $U_{\text{th}}=\sqrt{2\ln{K}}+O\left(1/\sqrt{\ln{K}}\right)$,
where $K=T\sqrt{\left|\frac{d^{2}C(0)}{dt^{2}}\right|}$ and $C(t)=\left\langle U\left(t^{\prime}\right)U\left(t^{\prime}+t\right)\right\rangle $
is the auto-correlation function of $U(t)$. These results are valid
provided that $C(t)$ decays to zero at long times and $K$ is large
\footnote{See EPAPS Document No. {[}{]} for Supplementary Material.%
}. Note that for a kernel $u(t)$ in eq. \eqref{eq:Tempotron_voltage-1}
of the form of difference of exponentials, $K$ takes the value of
eq. \eqref{eq:K}. 

\begin{figure}
 \includegraphics[width=1\columnwidth,height=0.4\columnwidth]{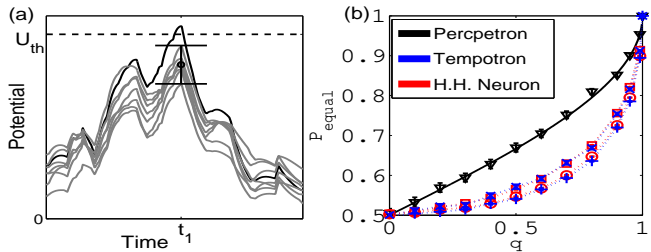}
\caption{\label{fig:P}(a) Time traces of the potentials of a Tempotron with
random weights, $U_{1}(t)$ (bold line), and of seven other Tempotrons,
$U_{2}(t)$ (gray lines), having overlap $q=.8$ with the first one.
The pattern is the same for all Tempotrons, and is classified as $+1$
by the first Tempotron: $U_{1}\left(t\right)$ is maximal in $t_{1}$,
and exceeds $U_{\text{th}}$. The error bar is centered in $t=t_{1},\bar{U}_{2}=q\, U_{1}\left(t_{1}\right)$,
and has height $\sqrt{1-q^{2}}$. Parameters are $K=100,\ N=1000$.
(b) Probability that 2 neurons will classify a random pattern in the
same manner, $P_{\text{equal}}$, vs. overlap between their weight
vectors, $q$, for the Perceptron (theory and simulations in black),
the Tempotron (blue $\times$ and $+$ symbols correspond, respectively,
to $K=100$ and $K^{\prime}=4K=400$) and the Hodgkin-Huxley (red
squares and circles correspond to $T=1.5\ \text{sec}$ and $T^{\prime}=4T=6\ \text{sec}$
respectively {[}13{]}) models. }

\end{figure}

We now consider two such Tempotrons, with an overlap $q$ between
their two weight vectors. Let us choose a pattern that is classified
as $+1$ by the first and denote by $t_{1}$ the time at which its
potential reaches its maximum value $U_{1}>U_{\text{th}}$. Let us
denote the postsynaptic potential of the second Tempotron at time
$t_{1}$ by $U_{2}$. Conditioned on $U_{1}$, the probability distribution
of $U_{2}$ is Gaussian with mean $\overline{U}_{2}=q\, U_{1}$ and
standard deviation $\sigma=\sqrt{1-q^{2}}$. According to \eqref{eq:v_max}
$U_{1}$ is close to $U_{\text{th}}$, and we may approximate $U_{\text{th}}-\overline{U}_{2}\simeq(1-q)\sqrt{2\ln{K}}$.
Thus, as long as $1-q\gg\frac{1}{\ln K}$, the typical fluctuations
of $U_{2}$ which are of $O(\sigma)$ are much smaller than the gap
between $\overline{U}_{2}$ and the threshold (Fig. \ref{fig:P}a);
hence $U_{2}$ is very likely smaller than $U_{\text{th}}$. This
implies that the overall probability that the second Tempotron's potential
crosses the threshold at any time remains close to $1/2$, unless
\begin{equation}
q\geq1-O\left(\frac{1}{\ln K}\right)\ .\label{eq:q}\end{equation}
 Thus, two Tempotrons are likely to agree on their classifications
of a random pattern only if the overlap in their synaptic weights
is close to 1. This result is confirmed by the simulations shown in
Fig. \ref{fig:P}b. We also present the simulation results for the
Hodgkin Huxley model {[}13{]}, a classical biophysical model for spike
generation. Interestingly, despite its complex dynamics, the classification
pattern of a pair of Hodgkin-Huxley neurons is similar to that of
the Tempotron, indicating that this behavior does not depend on the
details of the spike generation but on the summation of input spikes
within temporal windows. In contrast, in the case of the Perceptron,
which lacks temporal windows, the probability that two weight vectors
agree on their classification increases roughly linearly with their
overlap, $q$ (Fig. \ref{fig:P}b). The above result provides a qualitative
explanation of the clustered nature of the solution space. Consider
one solution to the classification task. Very similar weight vectors,
with overlaps larger than $1-O\left(1/\ln K\right)$ are likely to
be solutions, too, and compose a very small connected cluster of solutions
around the first solution. On the other hand having any positive overlap
smaller than this scale, does not provide significant advantage in
terms of classification error. Hence, entropy pressure for decreasing
the overlap wins, yielding a vanishingly small overlap $q_{0}$ between
two typical solutions.

The fact that $q_{0}$ is small for all $\alpha$ has important consequences.
First, $q_{0}$ in general measures the strength of the correlations
between the solution weight vector and individual quenched learnt
patterns. Small $q_{0}$ implies, therefore, that the statistics of
the potential after learning is approximately Gaussian with variance
and mean which are governed by the requirement that random patterns
induce spiking with probability $\frac{1}{2}$. As described above,
this implies that the distribution of $U_{\text{max}}$ of learnt
patterns has a Gumbel shape. Furthermore EVT predicts that the number
of threshold crossings in a pattern of duration $T$, $N_{\text{spikes}}$,
obeys a Poisson distribution with a mean rate $r=\frac{\ln{2}}{T}$,
consistent with a $\frac{1}{2}$-probability of firing within time
$T$ \citep{leadbetter1983extremes}. These predictions are confirmed
by numerical simulations shown in Fig. \ref{fig:traces-1}b,c.

EVT provides a basis for estimating the value of the capacity. Drawing
on analogy from the replica calculations (\citep{monasson1994dsa}
and below), we estimate the entropy of clusters in the solution space,
$S_{cl}$, through $S_{cl}=(\ln{V}-\ln{V_{cl}})/N$, where $V$ and
$V_{cl}$ are, respectively, the total volume of solutions and the
typical volume of one cluster. As $q_{0}\simeq0$, $V$ is simply
the product of the probabilities that the Gaussian potential $U$
crosses the threshold for each $+1$ pattern and does not do so for
each $-1$ pattern: $V=\left(\frac{1}{2}\right)^{N\alpha}$. Assuming
that the typical cluster is of 'compact' shape, its volume is given
by $V_{cl}=(1-q_{1})^{N/2}$ where $q_{1}$ is the typical overlap
between solutions within the cluster and scales according to eq. \eqref{eq:q}
as $1-q_{1}=O(1/\ln K)$. We therefore obtain, \begin{equation}
S_{cl}\simeq\frac{1}{2}{\ln\ln K}-\alpha\,\ln2\ .\label{eq:sc}\end{equation}
 Classifications are possible as long as $S_{cl}>0$, which yields
the capacity \eqref{eq:alpha_c}.

The above results are supported by an independent statistical mechanical
study of a simpler model, the discrete Tempotron \citep[Sup. Mat.]{gutig2006tnl},
where time is discrete, $t=\ell\,\tau$, $\ell=1,2,3,...$, and the
potential $U_{\ell}$ is the sum of the synaptic weights $\omega_{i}$,
multiplied by the number of spikes emitted by input $i$ in the time-bin
$\ell$. The patterns to be classified are associated an internal
representation (IR), which consists of the set of time-bin indices
$\ell$ such that $U_{\ell}>U_{th}$. The weight vectors implementing
the same IR form a convex domain of solutions. As the entire solution
space is not expected to be convex, calculating its volume is a difficult
task. Instead, following \citep{monasson1994dsa,engel2001sml}, we
have calculated the average value of the logarithm of the number of
typical implementable IR domains, $S_{IR}$, as a function of $\alpha$.
The calculation, based on the replica method, involves two overlaps:
the intra-overlap of a domain, $q_{1}^{IR}$, and the inter-overlap
between two domains, $q_{0}^{IR}$. When $K=\frac{T}{\tau}\gg1$ and
$\alpha\gg\frac{1}{\ln K}$, we find $q_{0}^{IR}\sim\frac{\alpha}{\ln K}$,
$1-q_{1}^{IR}\sim\frac{1}{\alpha^{2}\ln K}$, and $S_{IR}$ given
by the right-hand side of \eqref{eq:sc}. Hence $q_{0}^{IR}$ vanishes
as long as $\alpha\ll\ln K$, and the scaling of $q_{1}^{IR}$ is
compatible with $q_{1}$ given by EVT. This calculation also enables
us to estimate the capacity at finite $K$ (See Fig. \ref{fig:capacity}a).
The similarity between quantities defined in terms of connected clusters
of solutions, and those defined in terms of IR domains is a consequence
of the binary character of the overlaps in the large $K$ limit. For
the same reason, further effects of replica symmetry breaking should
affect only subleading corrections to $\alpha_{c}$. Numerical simulations
show that the discrete Tempotron behaves very similarly to the continuous
time Tempotron (Data not shown). This implies that the computational
capability of the Tempotron is not sensitive to the detailed shape
of the temporal integration.

In conclusion, we have presented a theory of the computational capacity
of  a neuron that performs classification of inputs by integrating
incoming spikes in space and time and generates its decision via threshold
crossing. Importantly, the Tempotron is not constrained to fire at
a given time in response to a target pattern. Thus, by adjusting the
timing of its output spikes, the Tempotron can choose the spatio-temporal
features that will trigger its firing for each target pattern. Despite
the simplicity of its architecture and dynamics, this property of
the Tempotron decision rule yields a rather complex structure of the
solution space and accounts for the superior performances of the Tempotron
compared to the Perceptron and to Perceptron-based models for learning
temporal sequences \citep{bressloff1992perceptron} which specify
the desired times of the output spikes. 
\begin{acknowledgments}
We thank Robert G\"{u}tig for very helpful discussions. This work
was supported in part by the Chateaubriand fellowship, the Israel
Science Foundation, the Israeli Defense Ministry and the ANR 06 JCJC-051
grant. 
\end{acknowledgments}
\bibliographystyle{apsrev4-1}

\end{document}